\documentclass[sigconf, hidelinks]{acmart}
\usepackage{xspace}
\usepackage{enumitem}
\AtBeginDocument{%
  }

\acmISBN{978-1-4503-XXXX-X/2018/06}

\usepackage[most]{tcolorbox}

\definecolor{testColor}{HTML}{814AB7}
\definecolor{testBrown}{HTML}{6D3C1F}
\definecolor{white}{HTML}{FFFFFF}

\definecolor{purple5}{HTML}{7209B7}
\definecolor{purple4}{HTML}{580966}
\definecolor{purple3}{HTML}{770C8A}
\definecolor{purple2}{HTML}{9410AB}
\definecolor{purple1}{HTML}{C6259E}
\definecolor{pink}{HTML}{F83A90}
\definecolor{orange2}{HTML}{FA7543}
\definecolor{orange1}{HTML}{FC9D22}
\definecolor{yellow}{HTML}{FDC500}

\newtcolorbox{keyInsightsBox}{enhanced, opacityback=0.75,colback=purple1!25!white,width=0.48\textwidth, colframe=white,arc=6mm,boxsep=2pt,    %
  top=1pt,       %
  bottom=1pt}

\usepackage{xcolor}  %
\definecolor{darkblue}{RGB}{0, 0, 139}  %
\definecolor{black}{RGB}{0, 0, 0}  %
\newcommand{\pquote}[1]{\textit{``\textcolor{darkblue}{#1}''}}

\setcopyright{none} %
\renewcommand\footnotetextcopyrightpermission[1]{}
\usepackage{todonotes}

\usepackage{tabularx}
\usepackage{tabularray}
\usepackage{booktabs} %
\usepackage[table]{xcolor} %

\usepackage{epigraph}
\usepackage{framed}
\usepackage[most]{tcolorbox}

\begin{document}

\title[Product Manager Practices for Delegating Work to GenAI]{Product Manager Practices for Delegating Work to Generative AI: ``Accountability must not be delegated to non-human actors''}

\author{Mara Ulloa$^{1, 4}$, Jenna L. Butler$^{4}$, Sankeerti Haniyur$^{4}$, Courtney Miller$^{2, 4}$, Barrett Amos$^{4}$, \\Advait Sarkar$^{4}$, Margaret-Anne Storey$^{3, 4}$}
\affiliation{%
  \institution{$^{1}$Northwestern University, Evanston, IL, USA. Email: mara.ulloa@u.northwestern.edu \\
               $^{2}$Carnegie Mellon University, Pittsburgh, PA, USA. Email: cemiller@andrew.cmu.edu  \\
               $^{3}$University of Victoria, BC, CA. Email: mstorey@uvic.ca \\
               $^{4}$Microsoft, WA, USA \& Camb., UK. Email: jennbu, sahaniyur, barrett.amos, advait@microsoft.com  \\
               }
  \country{}
}

\renewcommand{\shortauthors}{Ulloa et al.}

\begin{abstract}
Generative AI (GenAI) is changing the nature of knowledge work, particularly for Product Managers (PMs) in software development teams. While much software engineering research has focused on developers' interactions with GenAI, there is less understanding of how the work of PMs is evolving due to GenAI. To address this gap, we conducted a mixed-methods study at Microsoft, a large, multinational software company: surveying 885 PMs, analyzing telemetry data for a subset of PMs (N=731), and interviewing a subset of 15 PMs. We contribute: (1) PMs' current GenAI adoption rates, uses cases, and perceived benefits and barriers and; (2) a framework capturing how PMs assess which tasks to delegate to GenAI; (3) PMs adaptation practices for integrating GenAI into their roles and perceptions of how their role is evolving. We end by discussing implications on the broader GenAI workflow adoption process and software development roles.
\end{abstract}

\maketitle

\section{Introduction}
Generative AI (GenAI) tools are rapidly reshaping software engineering (SE) workflows. Code generation systems such as GitHub Copilot and Gemini's Code Assist now enable developers to produce code faster than ever before \cite{GitHubCopilot, Google_GeminiCodeAssist, hellendoorn2021towards, tufano2022using, zhang2022coditt5, obrien2024prompt, ani2023recent}. Yet focusing solely on the developer's tasks risks overlooking the larger transformation underway on the end-to-end process of software creation which involves not just software engineers (SWEs), but also Product Managers (PMs). PMs are a specific type of knowledge worker who connect business goals, user needs, and technical execution in software development \cite{chisa2014evolution, Estes, ebert2014software}. Despite their pivotal role in shaping and ensuring successful software outcomes, PMs have not been studied much in the SE research literature \cite{Hyrynsalmi19}, and scholars have called for more research and frameworks to capture the nature of PM work \cite{Oruthotaarachchi25}. While there is a fast-growing body of work examining how SWEs use GenAI \cite{miller2025maybe, abrahao2025software, ahmed2025artificial, butler2025dear, obi2025identifying}, little is known about how GenAI is transforming the PM role. Moreover, this is an especially critical time for studying PMs: early evidence suggests that GenAI is reshaping roles and task distribution, with SWEs’ work shifting \cite{miller2025maybe, butler2025dear} and knowledge workers more broadly experiencing changes in both the perception and execution of work \cite{Law25, Kobiella24, Li24, Srishti24}.

Studying how PMs delegate tasks to GenAI offers valuable insight into how delegation in general arises, since the PM role positions them as delegators by design~\cite{Estes, strunk2024delegate}. Their role naturally requires balancing strategic oversight with effective distribution of responsibilities ~\cite{ebert07}, making them a compelling group to survey and interview in building an understanding of delegation values. Understanding how tasks delegation occurs within PMs workflows can inform delegation practices in broader GenAI knowledge work.

Recognizing this research gap in this moment of rapid technological change, we conducted a large-scale mixed-methods study with Microsoft PMs. As one of the world leaders in GenAI development, Microsoft operates at the frontier of adoption, with leadership actively encouraging use and employees gaining early access to tools, like GitHub Copilot and Microsoft Copilot. This distinct context provides a window into how GenAI may reshape the PM role more broadly as these tools and tools like them diffuse across the industry. Through our study, we surveyed 885 PMs and interviewed a subset of 15, exploring their usage patterns, levels of expertise, beliefs, barriers, and lessons learned. Through our research, we aim to answer the following questions:

\begin{itemize}
    \item \textbf{RQ1:} What are GenAI adoption rates, use cases, and perceived benefits and barriers for GenAI use among PMs? 
    \item \textbf{RQ2:} What values shape which tasks PMs delegate to GenAI?
    \item \textbf{RQ3:} How are PM delegation practices and the PM role itself evolving as GenAI becomes integrated into their workflows?
\end{itemize}

Our findings elucidate the overall state of PM GenAI use (RQ1, \hyperref[sec:framework]{Section \ref*{sec:landscape}}). We present a framework describing how PMs delegate tasks to GenAI at the individual level, while considering beliefs and expectations at the team and organizational levels (RQ2, \hyperref[sec:framework]{Section \ref*{sec:framework}}). 
Additionally, we report changes in the PM workflow, like how they view GenAI as a necessary new skill and how they adapt to using it, causing a perceived evolution of their role and the skillset needed to leverage GenAI in an impactful way(RQ3, (\hyperref[sec:practices_and_role_evolution]{Section \ref*{sec:practices_and_role_evolution}}). 

\section{Background and Related Work}

\subsection{The Software Product Management Role}
PMs are key internal stakeholders within software development teams. PMs generally follow either an individual contributor (IC) or managerial track (supervising others). ICs focus on specific product tasks, while managers oversee team performance and broader goals.

Across both IC and manager tracks,  responsibilities and expectations of PMs vary, often reflecting their diverse backgrounds and the needs of their organizations~\cite{kittlaus2017software, pattyn24}. Unlike their peers, such as SWEs who often have formal training in computer science (CS), PMs often come to their roles from a wider range of disciplines, e.g., business administration, psychology, and user experience~\cite{ebert2014software, kittlaus2017software}. 

PMs fill cross-functional gaps in SE teams, owning the product lifecycle from ideation through delivery and iteration~\cite{Estes, chisa2014evolution}. Blending technical and non-technical expertise, PMs act as liaisons who align products with user needs and market demands. To support their work, PMs use planning and evaluation tools such as roadmaps and business cases~\cite{ebert2014software}. Prior work has also examined how PMs balance ethical considerations with business goals and user satisfaction~\cite{Lev}. Estes et al. highlight PMs' essential competencies and their value within development teams~\cite{Estes}. 

While the software product management research area is relatively young~\cite{Hyrynsalmi19}, some research describes: the role of PM in software startups~\cite{pattyn24}; challenges PMs encounter~\cite{springer2022}; the specific activities and roles PMs fill~\cite{maglyas17, maglyas13}; the PM role in different environments~\cite{springer18}; and the impact PMs can have on SE overall~\cite{ebert07}.

Some studies have examined PMs’ use of GenAI within broader knowledge worker research. One study included three PMs alongside consultants, journalists, and graduate students, highlighting challenges in synthesizing unstructured data and calling for AI that supports collaborative transparency and reduces over-reliance~\cite{yun2025generative}. Other work explored how industry professionals, including PMs, engage in prompt engineering during GenAI prototyping~\cite{subramonyam2025prototyping}. More recently, studies have focused specifically on PMs, examining how they gatekeep responsible AI use~\cite{smith25} and manage AI-first products to drive key features~\cite{parikh25}.

Although recent work has explored how PMs use GenAI (e.g., gatekeeping responsible AI use or creating AI-first product~\cite{smith25, parikh25}), little is known about how PMs use AI more broadly or how they use it in AI-forward companies. Our study %
provides novel empirical insight into PMs’ engagement with GenAI and potential role evolution as these tools see wider adoption.

\subsection{GenAI Adoption and Delegation Practices}
\subsubsection{GenAI Adoption in knowledge work}
Knowledge workers represent a nebulous (and disputed) professional designation that includes workers in medicine, research, and engineering (e.g., SWEs and PMs)~\cite{cortada2009rise, machlup1962production}. The accelerated adoption of GenAI, exceeding that of both the personal computer and the internet, underscores its potential to fundamentally reshape and adapt knowledge work ~\cite{bick2024rapid}.

SWEs are integrating GenAI tools such as GitHub Copilot, Gemini, Goose, and GitLab Duo, into existing workflows to support various aspects of their coding process~\cite{GitHubCopilot, Google_GeminiCodeAssist, goose, gitlabduo, ani2023recent}. Moreover, SWEs are adopting GenAI to author code, understand large codebases, and conduct code testing and review contributions~\cite{ becker2025measuring, tufano2022using, zhang2022coditt5, obrien2024prompt, ani2023recent}.

While a growing body of literature highlights the potential benefits of adopting GenAI for knowledge work ~\cite{shamszare2023clinicians, baek2024researchagent, trinkenreich2025get, liang2024can, singh2023hide, weisz2025examining}, many scholars caution against adoption without careful consideration. They highlight GenAI's limitations: the potential for labor displacement and the devaluation of human expertise~\cite{bender2021dangers, Schroeder, Braun2025GenAi}. In SE, concerns have emerged around overstating GenAI's programming capabilities and the risk of lower-quality code, including the introduction of security vulnerabilities~\cite{khemka2024toward}. In response to these concerns, researchers have proposed frameworks for evaluating both the benefits and risks of GenAI integration~\cite{Storey2024Playbook, trinkenreich2025get} and have argued that assessments of GenAI should be grounded in real-world usage environments. Building on these insights, we examine PMs’ perceptions, adoption, barriers, use cases, workflow changes, and values guiding GenAI delegation.

\subsubsection{GenAI Delegation vs Usage}
Our study specifically aims to understand \textit{what values influence which tasks PMs delegate to GenAI}. This understanding in turn can be used to inform organizational practices, and potentially the design of GenAI tools for PMs.

It is important to distinguish between \emph{delegation} and \emph{usage}. \emph{\textbf{Delegation}} occurs when a worker decides \textit{``to transfer rights and responsibilities for execution of a task or decision-making and the associated outcomes''}~\cite{strunk2024delegate} to an AI system—prior to use. In contrast, \emph{\textbf{usage}} is the act of applying GenAI to a task after such a decision has been made. A recent systematic review highlights that human-agent delegation research remains in its early stages~\cite{strunk2024delegate}.

Research has started mapping how workers use or envision GenAI, including creating, searching, manipulating information, seeking advice, or integrating AI with minimal context, e.g., in SE, LLMs can infer if code review comments match pull request changes. ~\cite{BrachmanMichelle, brachman2025current, hellendoorn2021towards, tufano2022using, zhang2022coditt5}. Beyond capturing GenAI usage patterns, it is also key to understand how knowledge workers reason about \emph{which tasks} to delegate, something we aim to address. 

The decision to delegate a task to GenAI—which precedes actual usage—involves a complex judgment process. Decision-making typically requires methodically narrowing options, weighing benefits and drawbacks, and arriving at an intentional choice \cite{jonassen2012designing, bruch2017decision}. Deciding whether to delegate or use GenAI is a nuanced form of problem-solving: workers must navigate the \emph{``jagged technological frontier''}, where GenAI performs some tasks easily while struggling with others of comparable complexity~\cite{dell2023navigating}.

Prior work described how knowledge workers use GenAI~\cite{BrachmanMichelle, brachman2025current}, and other scholars have highlighted that usage also involves emotional, metacognitive, and social dynamics ~\cite{tankelevitch2024metacognitive, sarkar2023exploring}. These dynamics often exert their influence during the deliberation and delegation phase that precedes actual usage. However, we lack an empirical basis for reasoning about how these dynamics influence delegation to GenAI for the PM role. We address this gap with our study, deriving a framework for selective delegation (\hyperref[sec:framework]{Section \ref*{sec:framework}}).

\section{Methods}
We used a concurrent mixed methods approach. Across Microsoft, we conducted a survey and in parallel conducted one-on-one semi-structured interviews. The survey allowed us to understand GenAI adoption patterns at scale, while the interviews allowed us to probe deeper into PMs’ GenAI usage, motivations, concerns, and workflows. We collected telemetry data from consenting survey participants to confirm self-reported GenAI usage and to supplement demographic information. Our study follows best practices for mixed-methods empirical SE research ~\cite{storey2025guiding} and was approved by Microsoft's IRB. The study was conducted from April to June 2025.

\subsection{GenAI-Early Adopter PM Environment}
We conducted this study at Microsoft, a large, multi-national software company with over 100,000 employees. Microsoft is an ``AI forward'' company, an early adopter of and advocate for GenAI tools. Microsoft also designs GenAI tools, provides licenses for its employees and usage is encouraged company-wide. This licensed access had been in place for over three years during our study. Our population included both Individual Contributors (ICs), who handle day-to-day PM work, and People Managers, who combine PM duties with managerial tasks (planning, managing others) and thus do tasks like performance reviews (hence referred to as ``Managers'').

\subsection{Instrument Design and Data Collection}

\subsubsection{Survey Design, Recruitment, and Data Collection}
To design our survey, we first conducted informal interviews with ten PMs, recruited through convenience sampling. These interviews helped us understand cultural and contextual factors, e.g., common jargon, challenges, and varying levels of GenAI use-- informing the key topics, concerns, and values mirrored in our survey questions.

Since managers and ICs have different perspectives, we branched the survey based on role.\footnote{\href{https://docs.google.com/document/d/1tuNQB6ij9wxsO-6Rmd2Mn4RiLPy3YgNl6Sw2vlRdGl8/edit?usp=sharin}{This is a link to the survey, interview guide, and a methods diagram}} \textbf{IC questions} focused on daily usage and beliefs and questions on four categories: Current GenAI Usage (5 questions), Attitudes \& Perceptions (5 questions), Barriers \& Opportunities(7 questions), Future Outlook \& Feature Requests (6 questions). \textbf{Manager question}s focused on observed team dynamics and organizational strategies, specifically the areas of Team GenAI Adoption \& Barriers (5 questions), Managerial Responsibilities \& Decision-Making (5 questions), Future Outlook for Team (3 questions), Leadership \& Culture (6 questions). Additionally, both groups were asked about challenges: e.g., ICs were asked \textit{``Is there a technical or workflow limitation that makes using GenAI difficult for you''} while managers were asked \textit{``What are the biggest challenges you and your direct reports face when trying to use GenAI for your team?''}

We piloted the survey with the PM team in our company division (44 ICs and manager PMs). Feedback included moving demographics to the end to emphasize core questions, grouping Likert items into blocks, and making minor wording edits. We also shared the survey with four Microsoft SE researchers (none authors in this study); one suggested adding two Work Trend Index questions~\cite{WorkTrend2025}—on Copilot as a thought partner and as a tool—both were added to the survey. We emailed the survey to 5,111 PMs in one Microsoft division. Over a 21-day period, 885 PMs (17\%) took an average of 13 minutes to complete it (\autoref{table:combined_demographics} shows demographics for 731 PMs consented for us to examine their Telemetry data).

\subsubsection{Interview Design, Recruitment, and Data Collection}
In parallel to running the survey, we conducted 15 semi-structured interviews with early consenting survey respondents. We applied purposive sampling to ensure diversity across geography, GenAI usage levels, job level, education, and gender (\autoref{table:combined_demographics}). Interviews lasted on average 30 minutes, guided by a semi-structured script covering role, workflow, GenAI usage, barriers, and future needs. We conducted the interviews and generated automatic transcriptions using Microsoft Teams.

\subsubsection{Telemetry Data Collection}
We collected telemetry data for survey respondents who consented to share it. This allowed us to keep the survey demographic section brief, as telemetry captures: the (1) individual's title, and (2) country the individual worked in \autoref{table:combined_demographics}. The telemetry data also included (3) frequency of interaction with GenAI tools. In the survey we asked participants to self-report, ``\textit{How often do you use Generative AI tools for your PM tasks?}'', allowing us to compare self-reported usage to actual usage.

\subsection{Data Analysis}
\subsubsection{Qualitative Data:} 
We analyzed the interviews before fully analyzing survey responses. To analyze the qualitative data from 15 semi-structured interviews and open-ended survey responses, we used reflexive thematic analysis~\cite{braun2006using}. We opted for this interpretive approach as our goal was to generate rich, nuanced insights rather than quantify responses. In reflexive thematic analysis, researcher subjectivity and reflexivity are considered integral to theme development, making inter-rater reliability metrics incompatible with the method~\cite{braun2006using, braun2019reflecting}. Instead, we ensured rigor through consensus validation: codes were collaboratively reviewed and refined during bi-weekly team meetings until verbal agreement was reached. This iterative process prioritized depth of understanding and reflexivity over statistical measures of agreement.

For interview analysis, one author, who also conducted the interviews and is trained in qualitative analysis, began by randomly selecting five transcripts to ensure variation and performed open coding while reviewing transcripts for accuracy and summarizing key points. This inductive approach allowed salient participant perspectives to emerge without imposing a priori categories. The same author then used affinity diagramming to cluster related observations and document emerging patterns and analytical insights~\cite{miles1994qualitative, beyer1999contextual}, producing a preliminary codebook of high-level codes. All authors reviewed and refined this codebook through bi-weekly discussions, reflecting on positionality, assumptions, and biases, and aligning themes with study objectives (e.g., usage patterns, delegation behaviors, role expectations). Codes offering novel insights relative to prior literature were prioritized. Using the evolving codebook as a guide, the lead researcher coded the remaining interviews, adding new codes as needed. We iteratively refined the codebook until no new codes emerged, indicating saturation. %

For survey responses, three authors each selected one open-ended question and independently performed open coding for all responses. The team then met to review initial codes, resolve discrepancies, and consolidate overlapping codes. Three questions were double-coded by two authors; the remaining questions were coded by one researcher. All codes were subsequently reviewed in team meetings, where disagreements were discussed until consensus was reached. This process produced a shared codebook that integrated insights from both interviews and surveys.

\subsubsection{Quantitative Data: Statistical Analysis of Survey and Telemetry Data}
The survey included both quantitative and qualitative items. For quantitative responses (e.g., Likert-scale, multiple-choice), we computed descriptive statistics (mean, standard deviation) and applied inferential tests (ANOVA, Mann–Whitney U) to examine relationships such as self-reported versus actual GenAI usage as indicated from the telemetry data. Analyses were conducted using Visual Studio, Jupyter Notebook, and Python scripts.

\subsubsection{Integrating Survey and Interview Data: Exploratory Sequential Analysis}
Although interviews and surveys were conducted in parallel, we analyzed interviews first to help us interpret survey findings by revealing the context behind participants’ responses. Insights from the interviews informed our coding of open-ended survey questions and guided interpretation of quantitative patterns. For example, our survey data indicated that many participants first engaged with GenAI through self-exploration, which aligned with interview themes emphasizing personal experimentation as a source of value. Conversely, divergence between data sources %
helped contextualize and deepen our understanding of participants’ perspectives.
For example, the survey showed some managers were not concerned about job displacement, but our interviews revealed this was due to impending retirement. While not concerned for themselves, they expressed concern for their teams and emphasized preparing them with skills to avoid risks from GenAI use.

\begin{table}[ht]
\scriptsize
\centering
\begin{tabularx}{\columnwidth}{lXrr}
\toprule
\textbf{Category} & \textbf{Characteristic} & \textbf{Survey (n=731)} & \textbf{Interview (n=15)} \\
\midrule\midrule
\textbf{Role} 
  & \hspace{1em}Individual Contributor & 604 (83\%) & 12 \\
  & \hspace{1em}People Manager         & 127 (17\%) & 3 \\
\midrule\midrule
\textbf{\shortstack[l]{YoE}}
  & \hspace{1em}< 1 year     & 28 (4\%)  & 1 \\
  & \hspace{1em}1--3 years   & 125 (17\%) & 4 \\
  & \hspace{1em}4--6 years   & 186 (25\%) & 1 \\
  & \hspace{1em}7--10 years  & 114 (16\%) & 4 \\
  & \hspace{1em}10+ years    & 278 (38\%) & 5 \\
\midrule\midrule
\textbf{Gender} 
  & \hspace{1em}Male                 & 436 (60\%) & 10 \\
  & \hspace{1em}Female               & 270 (37\%) & 4 \\
  & \hspace{1em}Prefer not to say    & 17 (2\%)   & -- \\
  & \hspace{1em}Non-binary           & 2 (<1\%)   & -- \\
  & \hspace{1em}Queer/Non-conforming & 1 (<1\%)   & 1 \\
\midrule\midrule
\textbf{Country} 
  & \hspace{1em}USA        & 474 (66\%) & 7 \\
  & \hspace{1em}India      & 58 (8\%)   & -- \\
  & \hspace{1em}Ireland    & 26 (4\%)   & 1 \\
  & \hspace{1em}Canada     & 19 (3\%)   & 3 \\
  & \hspace{1em}Other      & 135 (21\%) & 4 \\
\midrule\midrule
\textbf{Education} 
  & \hspace{1em}Computer Science   & 379 (30\%) & 7 \\
  & \hspace{1em}Engineering        & 263 (21\%) & 2 \\
  & \hspace{1em}Business Admin.    & 224 (18\%) & 5 \\
  & \hspace{1em}Economics          & 69 (6\%)  & 2 \\
  & \hspace{1em}Other               & 314 (25\%) & 10 \\
\bottomrule
\end{tabularx}
\caption{Demographics of the 731 survey respondents (who consented to sharing telemetry) and 15 interview participants.  Note some participants reported more than one educational background. YoE = Years of Experience}
\label{table:combined_demographics}
\end{table}

\section{Findings}

\begin{figure}
  \includegraphics[width=\linewidth]{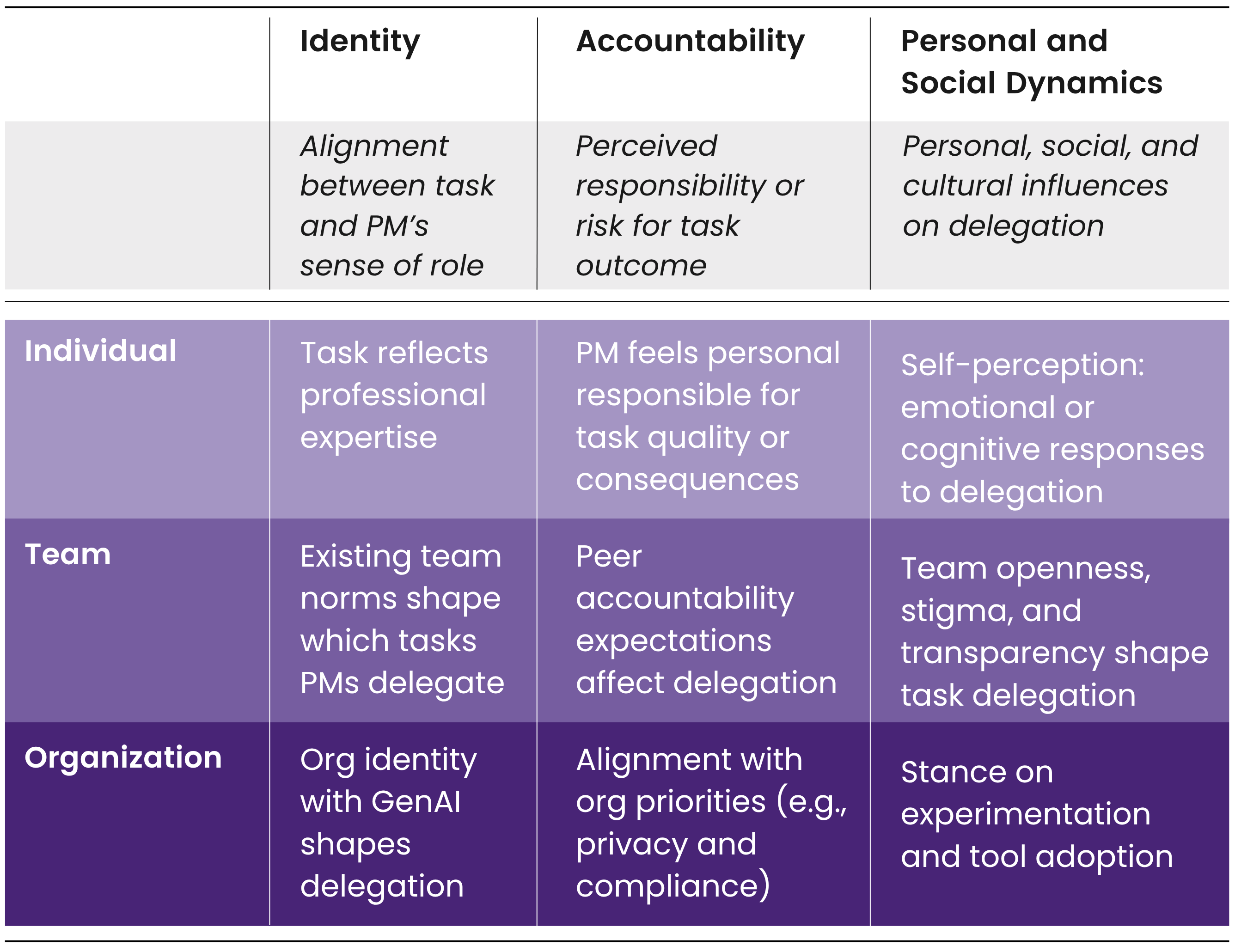}
  \caption{Product managers (PMs) identified values at the individual, team, and organizational levels influencing task delegation to GenAI-driven tools. These are discussed further in Findings, \hyperref[sec:framework]{Section \ref*{sec:framework}}.}
  \Description{}
  \label{fig:framework}
\end{figure}

We present findings across our three research questions. \textbf{RQ1}: What are GenAI adoption rates, use cases, and perceived benefits and barriers for GenAI use among PMs? (\hyperref[sec:landscape]{Section \ref*{sec:landscape}}), sets the stage by describing how PMs currently engage with GenAI and what they believe about its use. Building on this foundation, through \textbf{RQ2}: What values shape which tasks PMs delegate to GenAI? (\hyperref[sec:framework]{Section \ref*{sec:framework}}), we identify the value shaping delegation decisions. Finally, for \textbf{RQ3:} How are PM delegation practices and the PM role itself evolving as GenAI becomes integrated into their workflows? (\hyperref[sec:practices_and_role_evolution]{Section \ref*{sec:practices_and_role_evolution}}), we offer insights into how PMs develop the values depicted in the framework.\footnote{ Survey quotes are labeled PX, and interview quotes PXi, where X is the participant number.}

\subsection{PM GenAI adoption rates, use cases, and perceived benefits and challenges (RQ1)} \label{sec:landscape}
As little is currently known about how PMs use GenAI for their work, we started by trying to understand the current usage of GenAI by PMs including self reported usage amounts, what tasks they used it for, any benefits they find from using it, what challenges they experience in using it, and their overall beliefs about the tool. 

\subsubsection{Adoption and use cases}
A majority of ICs reported using GenAI daily or almost daily (62\%), with 16\% indicating usage 1–3 times per week, 3\% reporting usage 1–3 times per month, and less than 1\% reporting either never using it or using it less than once per month. To assess the accuracy of self-reported usage, we accessed participants’ M365 Copilot usage logs. We compared self-reported usage frequency with actual query counts using Spearman rank correlation, which yielded a positive correlation coefficient of 0.37. This suggests that higher self-reported usage is moderately associated with higher actual usage. Additionally, we conducted a one-way ANOVA to test whether mean query counts differed significantly across self-reported usage groups. The analysis produced an F-value of 6.58 with a p-value < 0.00001, indicating statistically significant differences in actual usage between the groups.

\begin{figure}
  \includegraphics[width=\linewidth]{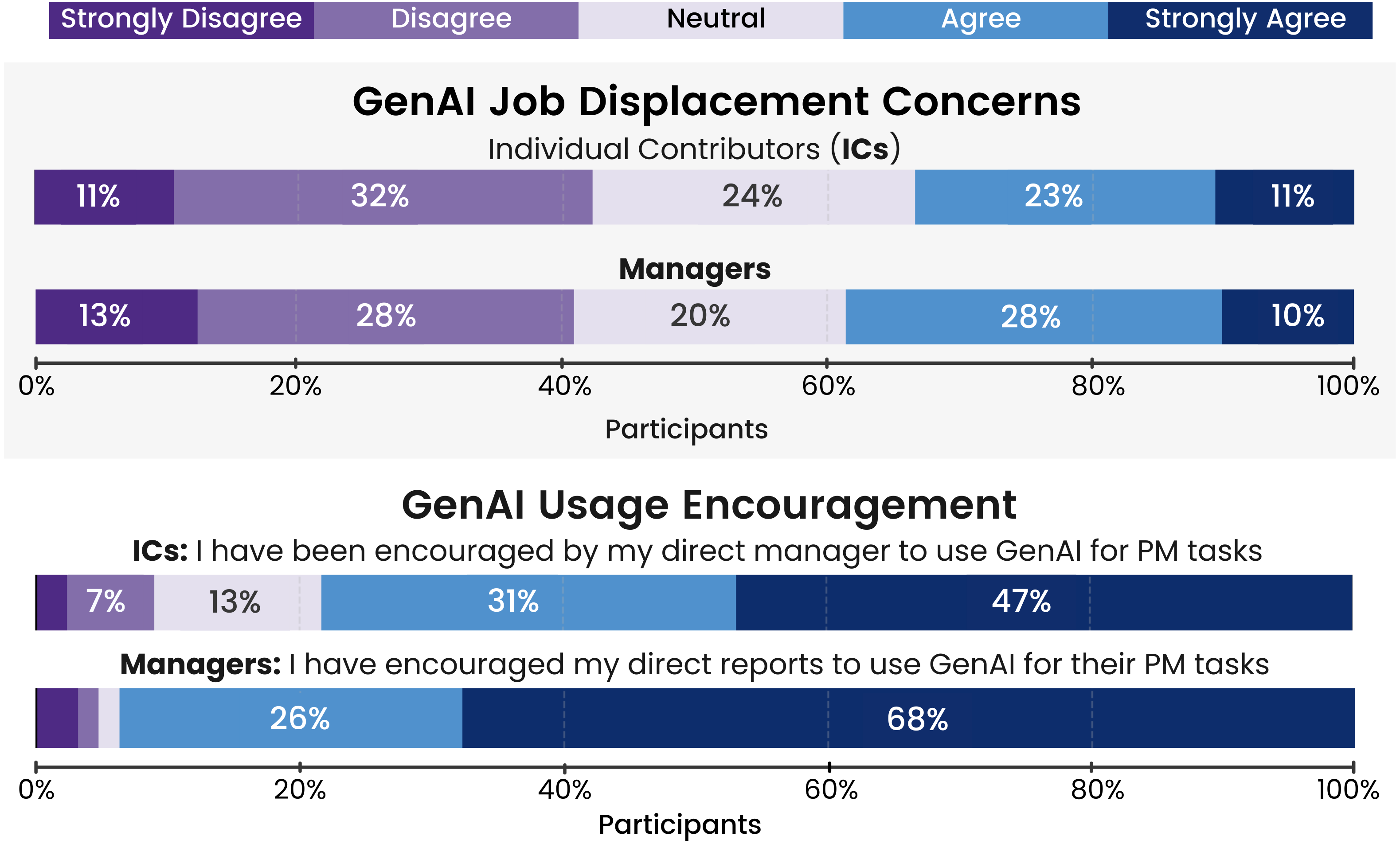}
  \caption{ICs and Managers shared how concerned they each were about GenAI job displacement. Additionally, high encouragement to use GenAI was reported by both groups.}
  \Description{}
  \label{fig:encouragement}
\end{figure}

We also examined whether participants were encouraged to use GenAI tools in their work. Forty-seven percent of individual contributors reported being encouraged by their direct manager to use GenAI, while 68\% of managers reported encouraging their direct reports to do so (\autoref{fig:encouragement}).

ICs also reported the \textbf{specific PM tasks} they use GenAI for with the majority of use cases involving documents and writing. 73\% of respondents reported using it for drafting or refining documents, 52\% use it for summarizing documents and 33\% use it for writing emails or external communications. Others used it to brainstorm ideas (35\%), conduct user research or data analysis (22\%), prototype ideas and write code (12\%), and learn a new skill (9\%).

For managers, we focused less on individual productivity and usage and instead asked for open text examples of how they have seen their team use it, as well as how they use it specifically for managerial tasks. 

We asked managers how they have used GenAI for \textbf{managerial tasks}; most reported using it to support performance evaluation of their employees in some way. Some reported that it helped them summarize peer feedback to share with their direct report (\pquote{I used it to organize [feedback] my directs received to identify trends in strengths and improvements.} P76); others shared that they use it as a way to potentially mitigate any bias they might have (\pquote{For [performance reviews], I use it to confirm delivery against core priorities to mitigate my own bias} P63). Some managers reported using it to help them ensure accuracy in their writing, such as P210 who reports, \pquote{Since both tone and wording significantly influence how feedback is perceived, I leverage LLMs to rephrase my feedback (direct or supportive), aligning it with my intended tone and goals.}

In addition to performance evaluation and communication, managers reported using it for team planning (\pquote{understanding and organizing team planning, drafting OKRs, allocating resources} P66), meeting management (\pquote{Organizing trainings sessions, including agenda and presentation. Organizing brainstorming and team building activities. [...] Taking meeting notes.} P852), and presentations and writing (\pquote{Rewriting summary mails/status updates and even generating a draft from scratch} P457).

\subsubsection{Perceived benefits of GenAI}
On a 5-point Likert scale, 81\% of ICs selected \textit{`Strongly Agree'} or \textit{`Agree'} in response to the statement \textit{Using GenAI often saves me time} (\autoref{fig:AI_attitudes}). Following this question, we had an optional open text question, \textit{In your own words, can you share a brief example of when GenAI saved you time or provided a clear benefit in your PM work?'}' These open text questions were coded and the most common response involved saving time in writing documents or presentations. One respondent said, \pquote{GenAI has helped me [turn] brainstorming notes into product spec requirements.  It has been able to draft problem statements and potential solutions based on my inputs.} (P84). The next most reported area of benefit was in summarizing and recapping meetings, as illustrated by P21, \pquote{Meeting recaps have been  incredibly useful, ability to draft comms or updates as well.} and P88, \pquote{GenAI recaps of Teams meetings and automated curation of action items is the best benefit I've experienced.} Some reported that prototyping was the biggest benefit, drastically reducing the time from idea to product, with P110 reporting, \pquote{Prototyping ideas is a lot quicker; I prototyped ideas on Failure Bucketing following a research paper. From paper to working prototype, it took me less than two hours. This would have taken 1-2 days of engineers' or my time to read and build a working prototype} %
Seven participants explicitly wrote that they could not provide an example of GenAI saving them time, such as P328 who reported, \pquote{It has not. Every time I've used GenAI it feels like I've asked an incompetent assistant to do my job badly.}

Again, with managers, we focused less on individual benefits and instead asked for examples of \textbf{how GenAI influenced a deliverable on their team}. Some reported a decrease in human effort needed for important tasks (\pquote{Time to closure and approvals has significantly decreased, and we have improved the auditability of our team's work without a lot of extra human effort.} P13), while others reported ways they had made their team more empowered and less dependent on themselves, as illustrated by P66 \pquote{I built an agent [...] focused on exec communication and attention, brevity, clarity -- I save a lot of time because I ask folks to run content through this tool before I review.} Some benefits were similar to those of ICs, such as improving writing and saving time with written documents: \pquote{By using GenAI, we were able to generate a very well written first draft of a [product release] blog post [...]. This is a key use case - AI is allowing us to go from 0-80\% on documents with just a few ideas and some links [...]. It prevents typical writers block situations [...] This allows us to advance ideas so much faster, universally accelerating our product thinking.} P142.

\subsubsection{Perceived barriers of using GenAI}
\begin{figure}
  \includegraphics[width=\linewidth]{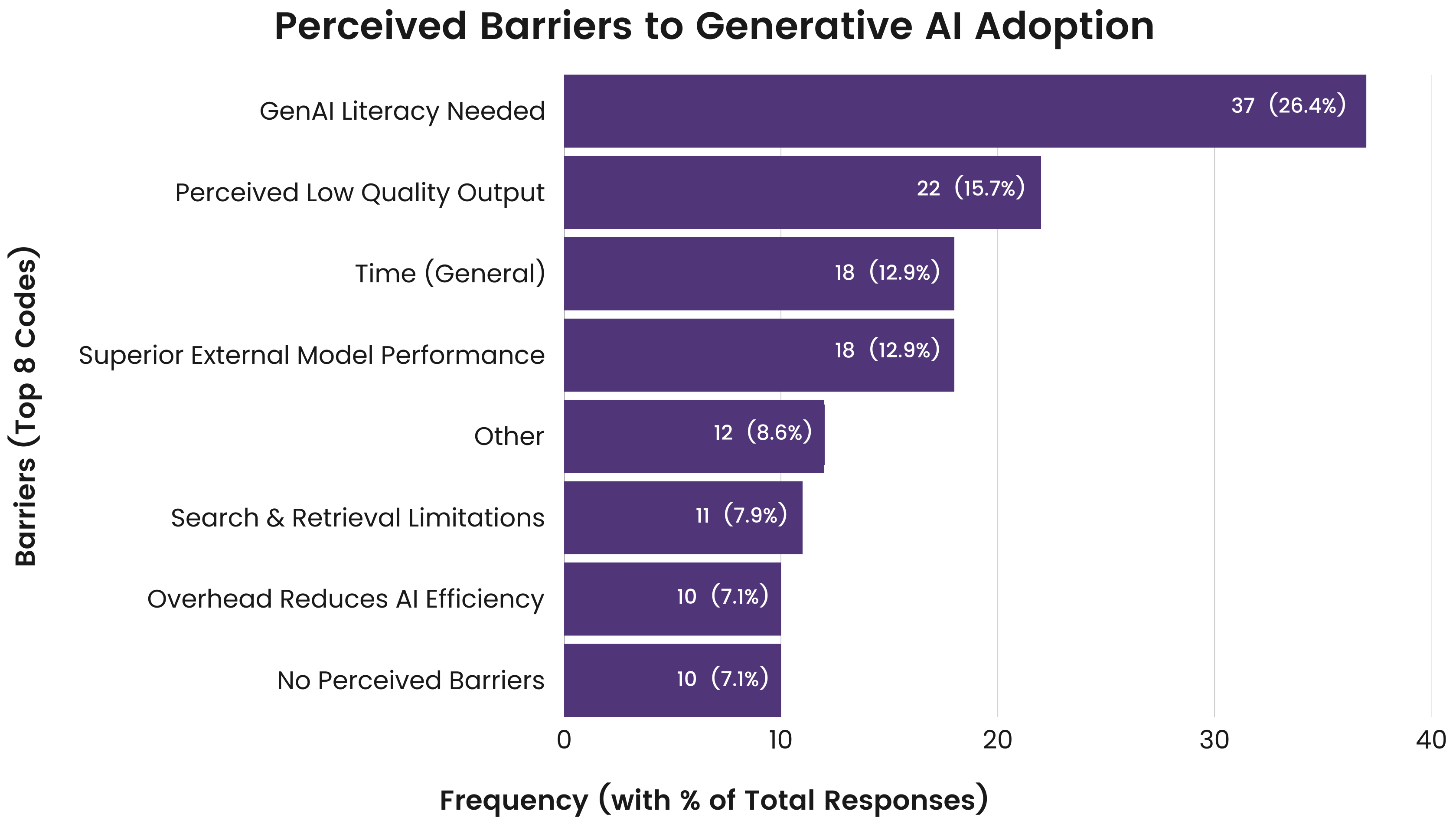}
  \caption{ICs' coded responses to the open-text question, \textit{Is there anything else you'd like to tell us about barriers to using GenAI more effectively in your role?}}
  \Description{}
  \label{fig:barriers}
\end{figure}

When asked what limits their GenAI use %
, one quarter selected “none,” indicating satisfaction with current usage. However, 47\% of IC PMs cited concerns about output quality or inaccuracies, and 37\% reported insufficient time to experiment. Other barriers included lack of relevant use cases (19\%), unclear data privacy/security guidelines (14\%), lack of knowledge or training (14\%), and unsatisfactory initial experiences leading them to abandon the tool (13\%).
We followed up with an optional open-ended question: \textit{“Is there anything else you'd like to tell us about \textbf{barriers to using GenAI} more effectively in your role?”} and coded 165 responses (\autoref{fig:barriers}). The most common theme was the lack of GenAI literacy, as P869 explained needing learning resources: \pquote{I need clear examples of what I can use GenAI for and how. Not to the PM role in general but to the implementation I have of the PM role.} Additionally, 22 participants said quality isn’t good enough: \pquote{Its accuracy (or lack thereof) is the single biggest barrier.} (P145). Other issues included lack of time to learn, perception that external models (unusable for work) are superior, and high overhead (e.g., validating output) making use impractical.

We also asked managers about issues their teams faced using a multi-select question. The most reported was \textit{difficulty integrating GenAI tools into existing workflows}, cited by 77 managers (51\% of 150). Next was \textit{not enough time to experiment / adopt new processes} (55\%), followed by \textit{concern about accuracy or bias} (54\%). Less common were lack of training (28\%), unclear company policy (28\%), and not knowing where to start (28\%). (As participants could select multiple options, totals exceed 100\%.)

\subsubsection{Beliefs and Feelings about GenAI}
\begin{figure}
  \includegraphics[width=\linewidth]{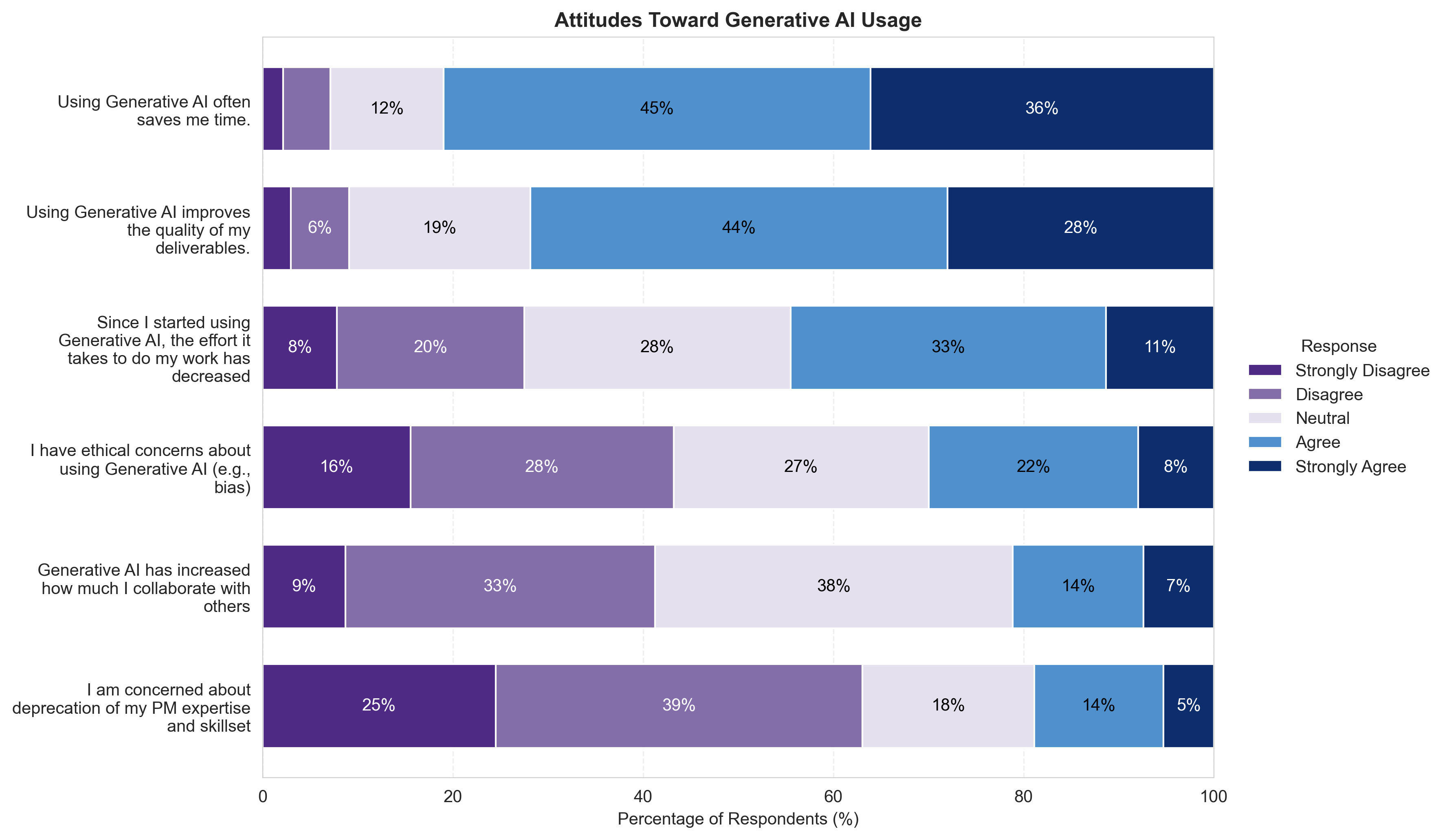}
  \caption{A selection of responses to Likert questions posed to IC PMs about their thoughts and beliefs on GenAI.}
  \Description{}
  \label{fig:AI_attitudes}
\end{figure}

Finally, we analyzed PMs perceptions and sentiments about this new way of working. We asked ICs a series of Likert questions to learn more about their views of GenAI, such as whether they trust it, if they think it improves work quality, and if they have ethical concerns (see \autoref{fig:AI_attitudes}). 

In addition to believing that GenAI saves them time, the majority of IC PMs (72\%) also believe that GenAI improves the quality of their work. However, the effort required to do their work hasn't decreased for most PMs (56\%). 67\% reported being aware of ethical considerations (such as bias, misinformation, etc) in these tools, but only 49\% said they regularly take steps to mitigate these issues. When asked if they have ethical concerns over the usage of GenAI, almost a third of IC PMs (30\%) responded that they do. When asked to elaborate on these concerns, people report being worried about not knowing how to handle bias (\pquote{I know that there are some biases in the GenAI but I wouldn't know what to do to course correct of fix it.} P26), worries about the environment, and concerns about deskilling, replacement or lack of growth opportunities (\pquote{I worry about becoming a data entry monkey with limited ability to grow. This feels like a tool that could easily remove the need for the middle structure of our company where people learn and grow in preparation for advancement all while it locks in top management and data entry /entry level personnel in their existing roles.} P355). Lastly, we also asked how they perceive GenAI as a whole. 69\% said they view Copilot as a tool that can help them and augment their work, while 38\% view it has a thought partner or teammate (some chose both conceptualizations). 

\subsection{Selective Delegation Framework: Deciding to delegate PM tasks to GenAI (RQ2)} \label{sec:framework}
Participants’ willingness to delegate tasks to GenAI was shaped by assessment of various values. Before using GenAI, PMs’ decisions were guided by: (1) alignment with their \textbf{identity} (who they are as a person), (2) a sense of \textbf{accountability} (what they or others expect of them), and (3) \textbf{personal \& social dynamics} (how they or others perceive them). Moreover, PMs described values influencing GenAI task delegation as operating at the individual, team, and organizational levels. We present a Selective Delegation Framework (\autoref{fig:framework})
to illustrate these relationships.

\subsubsection{Individual task delegation to GenAI}

Tasks that participants did not strongly \textbf{identify} with, such as formatting or summarizing meeting notes, were easily delegated to GenAI. However, when a task was central to their professional identity (e.g., writing for those with a literature background), participants were more reluctant to delegate to GenAI. For example, P815 shared, \pquote{I'm a journalist before this. I value my own voice and am not going to dilute it with AI, but also means it is not faster or more effective for me.} Similarly, P40i described written communication as core to the PM role, saying, \pquote{being a good communicator is important for everybody, but it's very important for PM. I take great pride on my written communication ability as a craft. If I let AI write something for me then they're no longer my words. It's not my craft anymore.}

This reluctance extended beyond writing to other tasks tied to PM identity. As P440 put it, \pquote{I'd like to see a shift in focusing on AI assistants that streamline menial tasks (task completion tracking, reminders, status updates), rather than our current focus on using GenAI to replace PM thought work (using GenAI for specs, strategies).} Participants felt GenAI should support, not replace, the cognitive and expressive aspects of their work. P365 emphasized this distinction, saying, \pquote{We need to create robots that do mundane tasks, not create systems that make us complacent and use less of our cognitive thinking and vocabulary.} Delegating high-identity tasks, for such participants, felt like losing a part of themselves.

Participants more readily delegated low-\textbf{accountability} tasks to GenAI, but hesitated with high-accountability work such as writing specifications or journey maps. As P661 said, \pquote{Accountability is one of the key aspects in a PM role. GenAI can [never] be used as a silver bullet because of this.} Even when using GenAI, PMs still felt accountable for outcomes. P547 shared, \pquote{As a PM, I used to write a newsletter myself. [However,] I have delegated this task to GenAI and it's now responsible but I am still accountable [...] it's posted under my name, any mistakes made by GenAI reflect poorly upon me.} Similarly, P805 noted, \pquote{Due to the nature of my job and the consequences \& risks of distributing incorrect information, I can't trust the outputs unless I have a way to quickly validate that the work done by AI is correct.} Participants emphasized the consequences tied to human accountability vs AI: \pquote{If a human REALLY messes something up... you go to jail. But what do you do when an AI really messes something up? It doesn't care.} (P40). Another participant, P640, explained why accountability must remain human-held \pquote{GenAI can augment the decision-making and... offer output to help decisions [but] \textbf{accountability must not be delegated to non-human actors}.} In essence, participants calibrated their use of GenAI to maintain a strong sense of accountability, even when delegating tasks.

Finally, PMs described a \textbf{personal dynamic} grounded in self-perception, reflecting on their emotional and cognitive responses when delegating tasks to GenAI. These reflections included emotions such as dissatisfaction, unease, increased productivity, or enhanced capability. For example, the thought of using AI alone elicited feelings of dissatisfaction: \pquote{When people suggest that I use AI to help me with my job, even for just drafting a document, I don't feel as satisfied as if I had done the thinking/writing/problem-solving on my own.} %
In contrast, some participants described GenAI as supportive, producing positive feelings such as reduced overwhelm and greater productivity: \pquote{As someone with a neurodivergent brain, GenAI has been beneficial in managing cognitive load and reducing overwhelm. [...] This support has been invaluable in maintaining clarity and focus, allowing me to leverage my strengths and contribute more effectively to my team.} (P368). Consistent with this perspective, another participant (P202) reported feeling more productive and found GenAI to be beneficial: \pquote{I feel even more productive, almost like a fellow team mate, a peer programmer is sitting next to me who is always there to work together.} Overall, PMs experienced a range of emotional reactions when using GenAI.

In summary, at the individual level, participants described a non-linear, three-step approach to GenAI. They considered their professional \textbf{identity} (\textit{Does using GenAI for this task encroach on my personal or professional identity?}). Next, they calibrated \textbf{accountability} (\textit{Is the GenAI accountable, or am I?}). Finally, they reflected on the emotional impact of the decision (\textit{How does this impact my self-perception}).

\subsubsection{Team level task delegation to GenAI}
In addition to individual values, PMs described their team as strongly influencing decisions to delegate tasks to GenAI. Values included existing team norms and identity, the need to maintain accountability, and the extent to which teammates supported GenAI usage.

Participants described how their \textbf{team’s established identity}, including norms that promote experimentation or maintain traditional non-GenAI practices, shaped their GenAI delegation decisions. For example, one participant stated \pquote{We are a data science team, so we've explored a lot of projects with GenAI as an outcome, but I haven't seen it change much about how we work at a team culture level. Innovation and experimentation was already the norm.} (P475) Another participant offered a similar perspective, noting that despite widespread use of meeting summarization and prompting across the organization, their team still expected manual note-taking: \pquote{My organization has leadership that has been around a long time and like things a certain way. For example, rather than just using Copilot for meeting notes they still expect notes to be captured and sent out the old fashioned way.} (P553). Regardless of the position of their teams on GenAI usage, participants emphasized that team norms consistently influenced their delegation behavior.

Participants identified \textbf{accountability} as a core PM skill requiring continuous effort. When delegating to GenAI, they considered its impact on team accountability. One participant suspected a teammate used GenAI without factoring in accountability: \pquote{I worry... it'll make people more complacent... I have worked with colleagues that have very clearly used GenAI without revising, and there is a lot of labor (on my part) involved in improving the quality of their contributions.} (P128). This increased their own workload and lowered team morale. Participants emphasized that ineffective delegation, without plans to preserve PM accountability, risked undermining customer-driven, data-informed product goals. Another participant noted that automation can erode essential skills: \pquote{I worry that we'll get to the point of AI talking to each other... and PMs no longer building the product judgment and core skills.} (P329). Overall, participants stressed delegating only when clear team strategies uphold accountability, ensuring GenAI does not degrade judgment or shared responsibility.

Participants described how \textbf{teammate dynamics} shaped their GenAI delegation decisions. Negative peer attitudes discouraged use, while support from peers and managers reinforced it. For example, strong anti-AI views from teammates made some hesitant: \pquote{There are 'anti-AI' folks withing my team, which makes me feel discouraged to use GenAI.} [P613] In contrast, recognition for AI use generated positive energy and encouragement: \pquote{The team, partnering with [other] teams, proposed a project during the last [day of learning], and won first place. This has created a positive energy among the team, and they to learn more about [GenAI].} (P204). Participants also described team strategies to fine-tune AI delegation and improve workflows: \pquote{My team has shared commitments in how we will a) advance our AI knowledge and industry awareness b) adopt AI tooling to improve our workflows / increase productivity.} (P825). Managers highlighted the combination of strategy and encouragement as a method for helping direct reports leverage GenAI: \pquote{Continue to encourage my team to learn and experiment, then share with the broader org what they are doing and how it's helping.} (P29). Overall, participants emphasized that social dynamics, whether judgment or support, significantly influenced delegation decisions.

\subsubsection{Organizational level task delegation to GenAI}
At the organizational level, participants highlighted the company’s identity as a GenAI leader, the importance of maintaining established organizational guiding principles, and a culture that strongly encourages GenAI use.

Participants often spoke proudly of their \textbf{organization’s identity} as \pquote{customer zero}, developing and actively using GenAI in-house. Regular engagement with GenAI was seen as essential for building cutting-edge tools for customers. For example, one participant encouraged: \pquote{Embrace being customer zero. Use the GenAI tools as frequently as possible. We are able to be the best technical trusted advisor when we can make the solutions relatable to a customer's mission, use cases, and pain points.} (P451). Others highlighted the many GenAI initiatives underway: \pquote{there are already soooooo many big initiatives here across the company.} (P28). A PM pursuing a Master's degree noted that while GenAI use was discouraged at university, it was strongly encouraged at their company: \pquote{For school, they don't want you to use it. But at the same time, everyone in industry is using it.} P04i. Overall, participants depicted their organization’s evolving identity as a key motivator for expanding GenAI adoption.

Organizational identity strongly shapes \textbf{social dynamics}. Employees are encouraged to use GenAI daily. ICs reported generally feeling supported, while 1 in 4 managers cited leadership pressure to drive adoption. One manager noted the future value justifies early use: \pquote{If AI isn't able to produce positive results, I encourage my team to try again in a few weeks. AI experiences and quality are increasing rapidly.} (P247). P871 observed that encouragement sometimes overlooked pitfalls: \pquote{We are not just encouraged but we have no choice but to use GenAI at work without discussing ethical concerns or possible implications to how it could affect our creativity and PM strengths.}.  Another participant remarked that the strong push might promote performative over utility: \pquote{There's too much of a visibility game going on where PMs are flooding channels with the slightest possible AI experience to appear relevant. The current environment makes folks afraid for their jobs, so we are getting more performative posturing than actual innovation.} (P59).

However, some participants felt that the AI push needed a more deliberate approach and should align with concrete use cases: \pquote{The use of AI should be handled as an Organizational Change Management initiative [...] While the technology is great, if an org or a person doesn't use it, it has no value to them.} (P299). Participants were generally receptive to organizational encouragement of GenAI but raised concerns about skill atrophy and stressed thoughtful integration into workflows.

As GenAI adoption grows, participants reflected on core organizational priorities. Many emphasized the need to uphold \textbf{longstanding values} such as security, privacy, and customer trust. One participant noted: \pquote{More focus and willingness to slow down in the name of privacy and security, even if that means we lose initial market leading status. Improve the models so the work actually gives us what we need and not wastes time.} (P345). A few participants expressed trust in the organization to safeguard data and interactions with proprietary GenAI tools: \pquote{I'm not super worried about AI risks such as privacy or content moderation issues. I feel Microsoft has solid privacy \& security controls around our AI tools and potential content issues just don't bother me at this point.} (P836). P828 also highlighted that the organization was integrating GenAI responsibly and hoped it would continue to mitigate risks: \pquote{We're doing a great job having a Responsible AI focus. I have done lots of work in this space, and trust and see we are driven by a great mission as a company, approaching this area with great care. At the same time, the risks in this space are unpredictable. This is the biggest change in tech in a couple of generations, maybe longer!}

\subsection{How PM delegation practices and role is evolving as GenAI is adopted (RQ3)} \label{sec:practices_and_role_evolution}

In \hyperref[sec:framework]{Section \ref*{sec:framework}} we present the Selective Delegation Framework describing how PMs decide to delegate tasks to GenAI. But \emph{how} did PMs develop this set of values? What processes might be causing not just individual delegation decisions in terms of their \textbf{identity} and \textbf{accountability}, but the \textbf{social dynamics} aspects of the delegation framework to evolve? 

To explore PM processes for the \emph{development of the skill of GenAI delegation}, we asked, \textit{If you could share one piece of advice with fellow PMs about using GenAI, what would it be?} Participants described two ways they evolve delegation: \textbf{direct experimentation to evaluate accountability}, and, for managers, \textbf{leading by example to navigate social dynamics}. They also saw a need to continuously adapt, viewing GenAI use as a \textbf{broadening of their PM identity} (e.g., engaging in technical work and blurring PM–SWE roles).

\subsubsection{Direct experimentation}
The most common advice from managers and ICs was to \textbf{“just start trying.”} Participants encouraged PMs to experiment without hesitation and to persist even when initial attempts fail. P275 emphasized \pquote{Use it to make you better / it's amazing} and P427 added, \pquote{Try it, try it again, and keep trying until you break it.} Beyond starting, PMs highlighted the value of returning to the tool and maintaining a growth mindset—focusing on learning rather than perfection: \pquote{Don't worry about not getting the prompt right the first time, just keep trying.} (P37) and \pquote{Jump in and try - don't let perfection be the enemy of good.} (P657). Finally, participants noted GenAI’s rapidly evolving nature, encouraging PMs to revisit it over time: \pquote{Keep trying it, it's always evolving so don't assume your experience today is the definitive way it will work a week from now. [...] whenever you're working on some task, think, is there some way I could leverage AI to make this better?} (P839).

In addition to starting with GenAI, PMs advised experimenting to understand its capabilities and discover what works best. P128i suggested, \pquote{Explore with it, play with it, and then celebrate the learnings.} Another participant, P10, added, \pquote{Use multiple GenAIs every day. Prompt every day. Evaluate what you liked/didn't like about the output. Compare them.} Some also recommended trying different tools and using experiments to guide future usage: \pquote{Continue using the tool to see how inputs produce certain outputs, learn how to dictate what you are looking for it to return to you.} (P91). Experimentation is key to refining GenAI delegation practices and understanding its evolving capabilities, which is essential for exercising judgement about \textbf{accountability} at individual, team, and organizational levels.

\subsubsection{Managers can serve as models of GenAI adoption for their team}
For managers, a key piece of advice was to \textbf{lead by example}. Some stated it plainly: \pquote{Be bold and lead with example.} (P55). Others offered more concrete guidance: \pquote{Lead by example, use it in your workflows and share the pros/cons with the team/learnings, etc.} (P76) Managers believe their own usage and enthusiasm will rub off, as P536 mentioned \pquote{Start with yourself. If the manager is using and talking about how helpful the tools are the teams will naturally try to help.} Finally, managers suggested sharing personal experiences can normalize experimentation and reinforce the importance of just starting: \pquote{Share what you've tried with your team. I use Copilot and other products at least once an hour. I share the successful, silly, confusing, etc results with my team to spark thinking for their own use cases and reinforce how easy it is to just jump in and try.} (P393). Just as direct experimentation allows PMs to develop and exercise judgments about accountability, managers attempting to model GenAI use allows individuals and teams to collectively develop their sense of judgments of \textbf{social dynamics}.

\subsubsection{PMs perceive the need for continuous personal improvement}

Many PMs acknowledged widespread concerns—among knowledge workers and in the media—about GenAI replacing jobs, but chose to focus on its potential benefits. One participant advised, \pquote{Don't be scared – it's an exciting time for technology – embrace AI, it's going to speed up a bunch of mundane tasks and give you the intelligence and a soundboard to discuss and sharpen your ideas.} (P651)

While replacement fears were less pronounced, participants emphasized the risk of falling behind. As one put it, \pquote{AI won't replace humans, but humans with AI will replace humans without AI.} (P376) A manager framed this similarly: \pquote{I firmly believe my staff won't be surpassed by AI, but we will be surpassed by someone who uses AI effectively.} (P879) Another participant echoed, \pquote{Someone not using AI can't compete with someone using AI, as someone with a 1990 flip phone can't compete with someone with an iPhone.} (P444) 

Several PMs described adapting to this shift as both natural to the PM role and essential for innovation. P833 reflected \pquote{It's a necessary area of upskilling to keep up with where the industry is going – the PM job isn't going away, but it's changing because of AI tools that will augment how we approach traditional tasks.} Similarly, P279 framed this evolution as core to PM identity: \pquote{A great PM should have the needed competencies to evolve in their role to what the upcoming roles that drive the creation of new technologies \& use of GenAI are.}

\subsubsection{Perceived blurring of PM and SWE roles}
A primary manifestation of this \textbf{shift in identity} is the blurring of PM and SWE roles. With GenAI, PMs felt capable of undertaking tasks traditionally performed by SWEs, such as writing scripts, developing prototypes, and creating agents. For example, P187 described how GenAI \pquote{helped me write SQL queries for data analysis that I would not be able to do myself}. In the survey, 91 (12\%) ICs reported using GenAI for prototyping and coding.

One manager shared using GenAI to prototype to bridge divides in expertise and communicate more effectively with developers, saying, \pquote{I've been using AI to come up with UI prototypes, see the concepts we want to get in front of users, improve language I use with developers.} (P13i) Another participant described feeling empowered because \pquote{PMs can take a more of a front-line role working with data and building prototypes, building agentic experiences.} (P66). 

Several PMs envisioned further blurring of traditional PM and SWE boundaries. P629 suggested, \pquote{it would be incredibly valuable to have sessions that help PMs get more familiar with dev-focused AI tools as well—especially since the division of responsibilities between PM and Engineering may evolve in this new AI-driven landscape.} While this requires stepping beyond the PM comfort zone, it offers an opportunity to develop new skills: \pquote{Get out of your comfort zone. Stop using AI to draft specs or to create messages. Try AI for areas that your peers work in (coding, designing, user research) and start learning how you can be great in some of those areas too. We all need to become more of a Swiss army knife.} (P236). Yet, rather than necessarily seeing this as broadening their official responsibility scope, some participants viewed this horizontal expertise as strengthening their core PM role: \pquote{I used GenAI during the planning process; I've been vibe coding to help me be stronger as a PM.} (P624).

\section{Discussion}
GenAI integration can reshape work by enhancing productivity, evolving roles, and enabling new human–computer interactions. Realizing this potential, however, requires a human-centered understanding of workers’ values and delegation practices. 

To examine PMs’ perspectives, we conducted a mixed-methods study with 885 survey responses, 15 interviews, and telemetry from 731 PMs. We analyzed (1) current GenAI usage and beliefs, (2) values shaping delegation, and (3) perceived adaptations to workflows and role evolution. Below we discuss a broader view of AI workflow adoption and (2) propose a rethinking of current software development roles.

\subsection{Broadening our understanding of AI workflow adoption}
We proposed a value-based Selective Delegation Framework (\hyperref[sec:framework]{Section \ref*{sec:framework}}) describing PMs’ GenAI delegation behavior. PMs consider identity, accountability, and personal and social dynamics when deciding which tasks to delegate or retain. The framework builds on prior research on AI workflow adoption across knowledge work~\cite{ulloa2022invisible, tankelevitch2024metacognitive, sarkar2023exploring, sarkar2024ai, drosos2024s}, showing that integration challenges often arise from social dynamics rather than technical gaps, emphasizing the need for sociotechnical approaches overly purely algorithmic optimization.

Previous work in AI workflow adoption has focused on the types of tasks individuals use AI for: creation, information seeking, or advice support ~\cite{BrachmanMichelle, brachman2025current}. The Selective Delegation Framework complements this work by furthering our understanding of how individuals first decide to delegate tasks to GenAI. \emph{Delegation} can be understood as a process of deciding to entrust a task to a technological system prior to usage. In contrast, \emph{usage} is the action taken after a decision to delegate has been reached.

Similarly, the Selective Delegation Framework extends prior work on decisions knowledge workers make about AI adoption. For instance, work on GenAI's metacognitive demands (monitoring and controlling one's own thought processes) describes the challenge of knowing whether, when, and how to incorporate GenAI into workflows as a worker's \emph{`automation strategy'} ~\cite{tankelevitch2024metacognitive, sarkar2023exploring}. These prior studies observe knowledge work changing from content production to \emph{`critical integration'} of AI output into ones workflow. This integration requires the application of an essential human skill, critical evaluation. Rather than blanket adoption, PMs in this study performed \emph{`critical integration'}, assessing GenAI's usefulness for tasks, weighing its value and implications at multiple levels.

Furthermore, previous work has speculatively connected Argyris' \emph{`double loop model of how organizations learn from errors'} to models of GenAI workflow adoption ~\cite{sarkar2023exploring, sarkar2024ai, argyris1977double}. In the outer loop, workers go beyond simply modifying GenAI output to reflecting and making modifications to their workflows based on experience working with GenAI. However, such outer loop shifts have not been empirically documented in prior work. Our study demonstrates that, at least for the PM profession, GenAI adoption can result in broader adaptation of workflows. More plainly, PMs described both (1) workflow adaptations (delegation practices); and (2) an evolution towards learning skills atypical to their PM role (\hyperref[sec:practices_and_role_evolution]{Section \ref*{sec:practices_and_role_evolution}}).

Other work describes an \emph{`iterative goal satisfaction'} GenAI framework, where the user attempts to satisfy a series of goals in stages %
with GenAI assistance. While ~\citet{drosos2024s} outlined the \emph{'iterative goal satisfaction'} framework, they primarily focused on illustrating challenges at each stage via a short-term laboratory study. In contrast, we offer empirical evidence for the framework, as PMs demonstrated and discussed going through each stage, thereby extending their work and contributing concrete cases grounded in real-world PM practice.

While previous studies allude to a broader set of decisions around AI workflow adoption than those identified thus far, we have studied these decisions with a novel empirical basis, founded in the experiences of software PMs integrating GenAI at work, enhancing our understanding of the integration of GenAI into workflows.

Our findings suggest that plans to adapt workflows in response to GenAI ought to be staged, prioritizing a deep understanding of worker values and GenAI's limitations, so workers can iteratively build trust in delegating tasks and assess its long-term impact on work practices. %
In industry, purchasing GenAI licenses without evaluating workers’ situated practices reflects a short-sighted approach; treating AI as a ``magic bullet'' without a clear integration strategy poses a major challenge for executives ~\cite{xie2025exploring}.
 
Thus, while research notes a GenAI-driven shift from content creation to \textit{critical integration} and an inherently human ability to adapt (workflows), workers like PMs, intellectually creative by nature, are justifiably concerned about its impact on their roles and expertise identity. This makes it clear that GenAI is no panacea, and calls for further research aimed at anchoring GenAI to PM values and its limitations for PM work specifically, as we've seen with SWEs ~\cite{miller2025maybe, weisz2025examining}, extending the groundwork laid by our study.

\subsection{Rethinking Software Development Roles} 

Our findings reinforce a critical point: understanding the evolving PM role is essential for the SE community. Traditionally, PMs have been peripheral teammates in the software development lifecycle, e.g., translators between business needs and technical execution~\cite{ebert2014software}. With GenAI, this view may no longer strongly hold, as PMs are taking on technical tasks previously reserved for SWEs. PMs in our study reported experimenting with prototyping, querying, and completing tasks that used to be reserved for SWEs.

Although some PMs we surveyed engaged in SWE-like tasks, the PM role is highly variable, leading to diverse GenAI adoption challenges. While much research and training target code generation, PMs reported a major barrier: limited GenAI literacy tailored to their responsibilities (\autoref{fig:barriers}). They viewed GenAI literacy as essential for unlocking unrealized value from GenAI in their workflows. This gap between potential and realized value led to PMs hesitating and demonstrating only strategic adoption, as seen in the \textbf{Selective Delegation Framework}. Ironically, GenAI adoption, meant to empower developers, can create tension by encroaching on tasks they find meaningful. PMs note a similar impact of GenAI on their work, i.e., in our investigation GenAI was perceived as a threat to their professional \textbf{identity}. This growing tension between empowerment and the erosion of professional \textbf{identity} highlights the need for more intentional development, not just of GenAI tools, but also of roles, by facilitating clearer GenAI task \textbf{delegation} and thoughtful \textbf{workflow adaptations}. Recognizing early implications of role shifting and GenAI delegation behaviors, we call on both the PM's organization and the SE research community to critically reimagine the software development lifecycle, not by simply reshuffling tasks, but by rethinking traditional assumptions about \emph{how}, and by \emph{whom}, software is engineered. 

Specifically, we invite other researchers to investigate the impact of the blurred role demands we saw and what kinds of organizational support is needed. As one participant put it: \pquote{We need to invest more resources \& time into defining guardrails for each PM role \& task so creativity can flourish. Like a trapeze artist, we need the confidence that there’s a safety net underneath to encourage exploration of what’s possible.} (P679). Organizational support for sharing PM-focused GenAI use cases enables others to realize its value, while sustained use fosters understanding and self-efficacy \citep{kobiella2025efficiency}.

Finally, organizations mandating GenAI use \cite{Marr2025} can ease the burden on individuals by clarifying for PMs (1) \textbf{how their shifting role will be evaluated}, aligning professional growth with company goals, and (2) \textbf{how to mentor direct reports and early-career PMs}, strengthening leadership skills noted as a need in our study. Proactive efforts also help organizations anticipate shifts in SE \textbf{team social dynamics} and evolving notions of \textbf{accountability}, as P821 noted \pquote{The SWE role will move ‘up the stack’ to be a bit more high level. The PM job will move ‘down the stack’ and a bit closer to code. We will meet in the middle and swap a lot of accountabilities}.

\section{Limitations} 
As is inherent in scientific studies, our work carries some limitations. First, we conducted our investigation in a single company, which may limit generalizability. However, as other work notes, single case studies can lead to significant discoveries across other contexts \cite{flyvbjerg2006five}. We also highlight that this company’s global, multilingual teams make it more diverse than typical single-site research. Additionally, the culture at the company we recruited from strongly encourages GenAI usage; a study from a company where the inverse is true would yield a necessary complementary perspective.

Additionally, several limitations common to qualitative research may be embedded in our study. These include potential issues with question design quality~\cite{birmingham2003using}, the risk of researcher inexperience leading to overlooked insights~\cite{koskei2015role}, and the possibility of self-reporting bias during interviews~\cite{chalhoub2022data}. To mitigate these risks, all interviews were conducted and qualitative data was primarily analyzed by researchers trained in qualitative methods. Furthermore, we avoided making open-ended survey questions mandatory and posing leading questions to interview participants.

\section{Conclusion} 
GenAI is reshaping knowledge work, particularly for PMs—key contributors in SE teams and early adopters of these tools. Given their pivotal role, their experiences offer valuable guidance for designing GenAI-driven workplace tools. In this work, we presented PMs’ current GenAI usage and beliefs; the \hyperref[sec:framework]{Section \ref{sec:framework}} framework, which outlines how PMs consider identity, accountability, and both personal and social dynamics across individual, team, and organizational levels when delegating tasks to GenAI; and how they perceive workflow adaptations and role shifts. We discussed how our findings expand understanding of AI workflow adoption, its impact on software development, and provide recommendations for organizations and SE research. Our goal is to foster inclusive dialogue that empowers PMs to shape, rather than simply adapt to, GenAI workflows.

\bibliographystyle{ACM-Reference-Format}

\end{document}